\documentclass[superscriptaddress, aps, pre, twocolumn, floatfix ]{revtex4-2}

\usepackage{graphicx}
\usepackage{dcolumn}
\usepackage{bm}

\usepackage[utf8]{inputenc}
\usepackage[T1]{fontenc}
\usepackage[english]{babel}
\usepackage{graphicx}
\usepackage{epsfig}
\usepackage{epstopdf}
\usepackage{amsmath}
\usepackage{amssymb}
\usepackage{times}
\usepackage{setspace}
\usepackage{verbatim}
\usepackage{xcolor}
\usepackage{mathrsfs}
\usepackage{euscript}
\usepackage{mathptmx}

\usepackage{hyperref}
\hypersetup{
    colorlinks=true,
    linkcolor=blue,
    filecolor=magenta,      
    urlcolor=blue,
    citecolor=blue, 
    }

\usepackage[normalem]{ulem}

\begin{document}


\title{Unusual ergodic and chaotic properties of trapped hard rods}

\author{Debarshee Bagchi}
\email{debarshee.bagchi@icts.res.in}
\address{International Centre for Theoretical Sciences, Tata Institute of Fundamental Research, Bengaluru -- 560089, India}

\author{Jitendra Kethepalli}
\email{jitendra.kethepalli@icts.res.in}
\address{International Centre for Theoretical Sciences, Tata Institute of Fundamental Research, Bengaluru -- 560089, India}

\author{Vir B. Bulchandani}
\email{vbb@princeton.edu}
\address{Department of Physics, Princeton University, NJ 08544, USA}
\address{Princeton Center for Theoretical Science, Princeton University, NJ 08544, USA}

\author{Abhishek Dhar}
\email{abhishek.dhar@icts.res.in}
\address{International Centre for Theoretical Sciences, Tata Institute of Fundamental Research, Bengaluru -- 560089, India}

\author{David A. Huse}
\email{huse@princeton.edu}
\address{Department of Physics, Princeton University, NJ 08544, USA}

\author{Manas Kulkarni}
\email{manas.kulkarni@icts.res.in}
\address{International Centre for Theoretical Sciences, Tata Institute of Fundamental Research, Bengaluru -- 560089, India}

\author{Anupam Kundu}
\email{anupam.kundu@icts.res.in}
\address{International Centre for Theoretical Sciences, Tata Institute of Fundamental Research, Bengaluru -- 560089, India}

\date{\today}

\begin{abstract}
We investigate ergodicity, chaos and thermalization for a one-dimensional classical gas of hard rods confined to an external quadratic or quartic trap, which breaks microscopic integrability. To quantify the strength of chaos in this system, we compute its maximal Lyapunov exponent numerically. The approach to thermal equilibrium is studied by considering the time evolution of particle position and velocity distributions and comparing the late-time profiles with the Gibbs state. Remarkably, we find that quadratically trapped hard rods are highly non-ergodic and do not resemble a Gibbs state even at extremely long times, despite compelling evidence of chaos for four or more rods. On the other hand, our numerical results reveal that  hard rods in a quartic trap exhibit both chaos and thermalization, and equilibrate to a Gibbs state as expected for a nonintegrable many-body system.   
\end{abstract}

%

\maketitle


\section{Introduction}
\label{intro}
The question of how isolated many body systems thermalize is of long-standing interest; a canonical study is that of Fermi, Pasta, Ulam and Tsingou (FPUT)~\cite{fermi1955studies}. The surprising finding of FPUT was that a one-dimensional  anharmonic chain of oscillators did not exhibit equipartition of energy even at very long times, with the system showing quasi-periodic behaviour and near-perfect recurrences. Various mechanisms have been proposed to explain the results of FPUT~\cite{ford1992fermi, berman2005fermi, dauxois2005fermi, gallavotti2007fermi}, e.g, proximity to integrable models such as the Korteweg–De Vries equation~\cite{zabusky1965} or the Toda model~\cite{casetti1997fermi, benettin2013fermi, goldfriend2019equilibration} as formalized by KAM theory~\cite{henrici2008results}, the stochasticity threshold~\cite{israiljev1965statistical}, the presence of discrete breathers~\cite{flach2008discrete} and most recently the formalism of wave turbulence~\cite{onorato2015route}. 

One striking feature of this system is a separation between the timescales for equilibration and chaos. From numerical  simulations of the $\alpha$-FPUT model~\cite{casetti1997fermi,deluca1995energy, livi1985equipartition}, it was shown  that for generic initial conditions~\cite{casetti1997fermi}, the timescale for the system to thermalize (defined as the time to reach equipartition of energy) was much longer than the timescale needed to observe chaos (defined as the time for the system to escape from regular regions in phase space to chaotic ones), with both timescales increasing as the energy per particle decreased and appearing to diverge at some critical value. (However, recent studies based on wave turbulence seem to indicate the absence of any such threshold~\cite{onorato2015route}.) Some subtleties in defining thermalization times and their possible relation to Lyapunov exponents were investigated recently in Ref.~\onlinecite{ganapa2020thermalization}. Despite a vast body of literature on the topic, a definitive theory of thermalization in the FPUT chain still eludes the community and there appears to be no consensus.

More recently, another family of clean many-body systems that fail to thermalize under their own dynamics has been scrutinized. These systems consist of particles with integrable two-body interactions, which are placed in an external trapping potential that breaks both translation symmetry and integrability of their interactions. Given that the trap breaks integrability, such systems are na{\"i}vely expected to thermalize to the Gibbs ensemble, but a prominent experiment realizing a trapped Lieb-Liniger gas with ultracold rubidium atoms showed that this expectation was not warranted~\cite{kinoshita2006quantum}. A detailed theory of the resulting Newton's-cradle-like dynamics had to await the development of generalized hydrodynamics~\cite{castro2016emergent,bertini2016transport} (GHD). While the latter theory appears to be more than adequate for modelling short-time dynamics of such trapped integrable systems~\cite{doyon2017note,bulchandani2017solvable,cao2018incomplete,caux2019hydrodynamics,bastianello2020thermalization,bulchandani2021quasiparticle,malvania2021generalized}, the fate of these systems at long times and in the absence of experimental imperfections remains somewhat unclear.

For example, previous work on one-dimensional classical hard rods in an integrability-breaking quadratic potential~\cite{cao2018incomplete} found numerically that despite the dynamics exhibiting positive Lyapunov exponents, the system did not thermalize to a Gibbs state at the longest accessible simulation times (on the order of ten thousand periods of the trapping potential). Moreover, the long-time steady state was found to be a stationary state of the ballistic-scale (i.e. non-dissipative) GHD equations (as suggested previously~\cite{doyon2017note}). This observation, together with further numerical findings reported below, appears to be incompatible with a subsequent proposal~\cite{bastianello2020thermalization} that diffusive corrections~\cite{de2019diffusion,gopalakrishnan2018hydrodynamics} to generalized hydrodynamics inevitably lead to thermalization in integrability-breaking traps. Even if thermalization does occur for quadratically trapped hard rods at numerically inaccessibly long times, it remains to be explained why this timescale is so long. Systems of rational Calogero particles have also been found not to thermalize on accessible timescales in traps that are expected to break integrability~\cite{bulchandani2021quasiparticle} (though this is not in tension with theory~\cite{bastianello2020thermalization} insofar as diffusive corrections to the rational Calogero GHD vanish~\cite{bulchandani2021quasiparticle}). Finally, we note that the effect of integrability-breaking by traps has been studied for the classical Toda system~\cite{di2018transport,dhar2019transport}. In this case, it was found that the quadratically trapped system was weakly chaotic while the quartically trapped system displayed strong chaos and thermalization. 

Thus despite much recent progress, several fundamental questions concerning the thermalization of trapped integrable particles remain unresolved, including whether or not these systems are truly ergodic, whether they can support additional microscopic conservation laws, and how far these properties coexist with chaos. Another open question, in answer to which there is conflicting evidence in the literature, is whether the stationary state in a generic trap is the Gibbs state~\cite{bastianello2020thermalization} or one of infinitely many non-thermal stationary solutions to ballistic-scale GHD~\cite{doyon2017note,cao2018incomplete}. We will address some of these questions below.
 
In this paper, we study the effects of integrability breaking in one-dimensional systems of hard rods of length $a$ that are confined to external potentials of the form $U(x)=k x^\delta/\delta$ with strength $k>0$, where $\delta=2$ for quadratic trap and $\delta=4$ for the quartic trap. We diagnose chaos, ergodicity and thermalization in these systems through probes such as the maximal Lyapunov exponent (LE), equipartition of energy between rods, and the position and velocity distributions of the rods. We find that while quartically trapped rods behave like a typical non-integrable many-body system, quadratically trapped rods exhibit many drastically different and unexpected properties. The only additional microscopic conservation law for quadratically trapped rods beyond the total energy, $E$, appears to be the energy of the centre of mass, $E_{\rm cm} = \frac{1}{2N^2}\left[(\sum_i x_i)^2 + (\sum_i v_i)^2 \right]$. Nevertheless we find that the system appears to be non-ergodic, has unconventional chaos properties, and fails to thermalize to the Gibbs state even at extremely long times.

\begin{figure}[htb]
\hspace{-1.3cm} \includegraphics[scale=0.4]{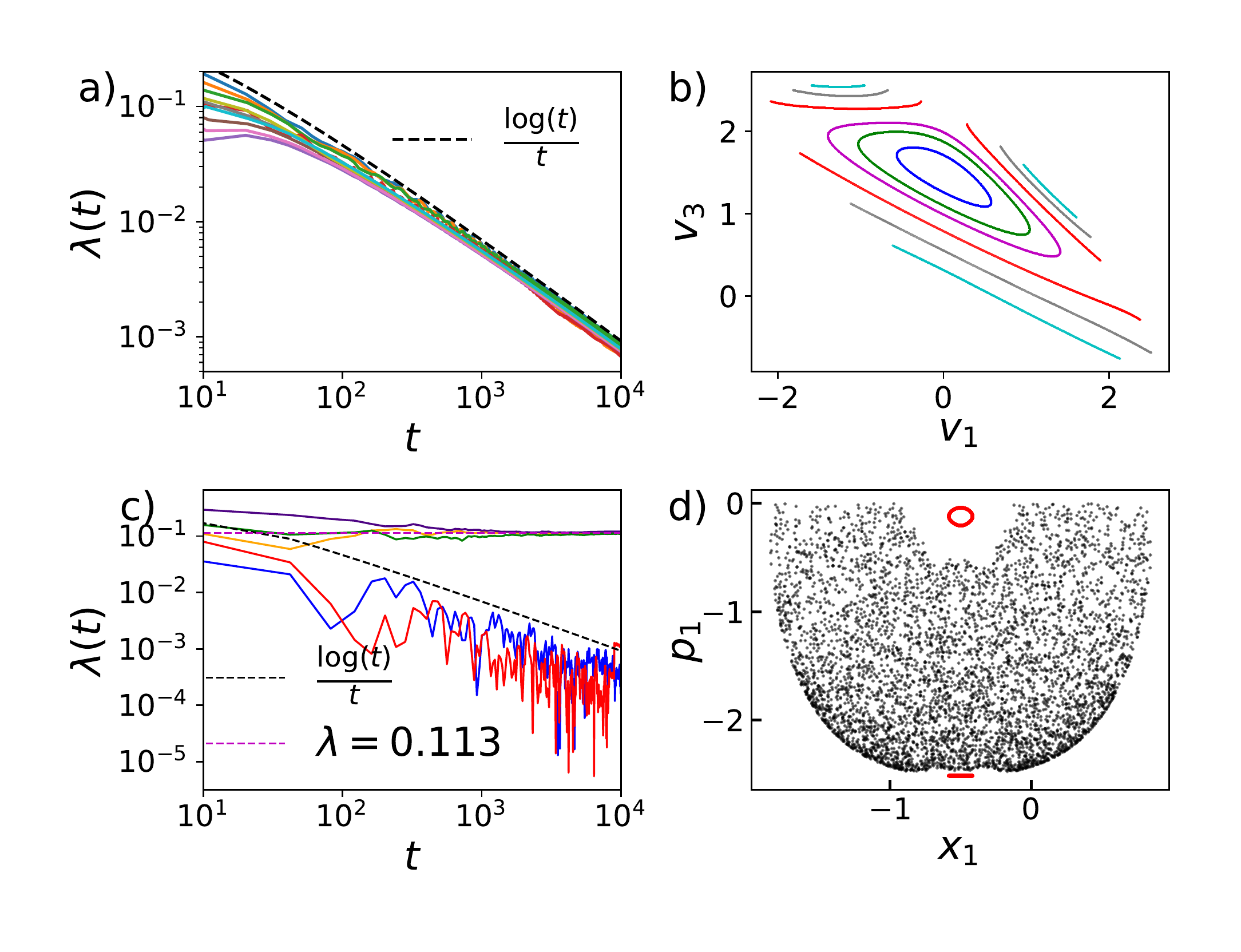}
 	\caption{ Plots of (a) time-dependent Lyapunov exponent $\lambda(t)$, and (b) Poincar\'e section for $N = 3$ rods in the quadratic trap for $10$ and $6$ different initial conditions,  respectively, with $E=6$ and $E_{\rm cm}=0$. Figures (c,d) show plots of $\lambda(t)$ and the Poincar\'e section for  $N = 2$ rods in the quartic potential for $5$ and $2$ initial conditions, respectively, with  energy $E=3.2$. To compute $\lambda(t)$, we used a linearized dynamics for (a)  and two trajectories with  $\epsilon=10^{-10}$ for (c). The $\log t/t$ behaviour in (a) and the regular sections in (b) are consistent with the integrability of $N=3$ rods in the quadratic trap. Interestingly, figure (c) reveals the existence of both chaotic and non-chaotic trajectories for the quartic case. This is also reflected in (d) where we observe two types of patterns, namely scattered (black) and regular (red).}
	\label{fig:LLExpN3}
\end{figure}

\begin{figure}[htb]
    {\includegraphics[scale=0.5]{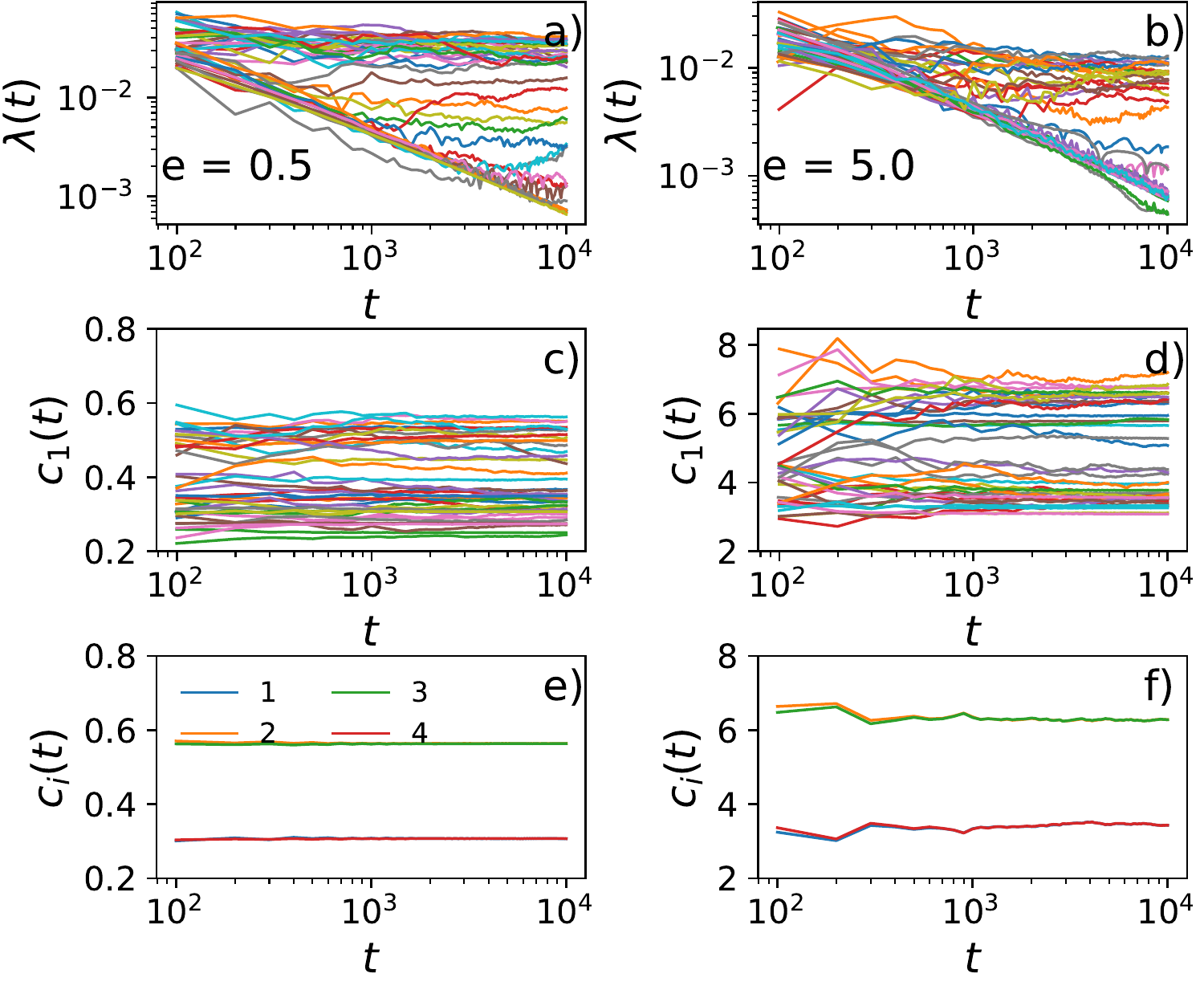}}
 \caption{Time evolution of $\lambda(t)$ and $c_1(t)$ for $N = 4$, starting from $40$ different initial conditions (each colour denotes one initial condition) generated using the SMED protocol, for (a,c,e) $ {\rm e} = 0.5$ and (b,d,f) ${\rm e} = 5.0$. This shows the strong dependence on initial conditions of the late-time values of $\lambda(t)$ and $c_1(t)$ for $N=4$ hard rods in a quadratic trap. In (e) and (f) we show the evolution of the time-averaged values of $c_i = T_i/N^2$ of $N=4$ hard-rods for ${\rm e} = 0.5$ and ${\rm e} = 5.0$ respectively, for one realization. In the long time limit, $c_1$ and $c_4$ are equal but have a value that is significantly different from $c_2$ and $c_3$, indicating a lack of energy equipartition.}
\label{fig:N4timeseries}
\end{figure}

Below we summarize our main findings (see Table.~\ref{tab_table1}):

\begin{enumerate}
    \item We find that in a quadratic trap, a system of $N=3$ hard rods shows a strong signature of integrability in the form of a vanishing maximal Lyapunov exponent (Fig.~\ref{fig:LLExpN3}a) and a regular Poincar\'e section (Fig.~\ref{fig:LLExpN3}b). This is in striking contrast to the case of two rods confined to a quartic trap, which has both finite (positive) and vanishing  Lyapunov exponents ( Fig.~\ref{fig:LLExpN3}c) and a mixed phase space with both chaotic and regular regions (Fig.~\ref{fig:LLExpN3}d). Our findings hint at the existence of more conserved quantities for three rods in a quadratic confining potential (see also \cite{cao2018incomplete}).
    \item For any finite number of rods $N>3$ in a quadratic potential, we find that the LE is positive. Nevertheless, we find compelling evidence that the system is highly non-ergodic. This is demonstrated by the strong initial-condition-dependence of the LE and the time-averaged kinetic temperature (Fig.~\ref{fig:N4timeseries}). Such non-ergodicity is further suggested by the broad distributions of Lyapunov exponents and rescaled temperatures (Fig.~\ref{fig:harmproblambda}). These distributions are obtained by time-evolving initial conditions that are sampled uniformly from the constant $E$, $E_{\rm cm}$ microcanonical surface (see Sec.~\ref{sec:num} for details). Remarkably, hard rods confined to a quartic trap exhibit qualitatively completely different behaviour, and we find evidence of conventional chaotic thermalizing dynamics expected for generic, non-integrable, classical many-body systems (Fig.~\ref{fig:quartproblambda}). 
    
    \item The system is described completely by two dimensionless  parameters: the rescaled energy, ${\rm e}=E/(N^{\delta+1} k a^{\delta})$ and the number of rods, $N$. For the quadratic case with fixed ${\rm e}$, the average maximal Lyapunov exponent $\langle \lambda \rangle$ converges to a finite value with increasing $N$. This converged value shows an $\sim {\rm e}^{-1/2}$ scaling over a wide range of ${\rm e}$ values~(Fig.~\ref{fig:LLExpc}a). In sharp contrast, for the quartic case the average LE ($\langle \lambda \rangle$) for a given $N$ grows as $\sim {\rm e}^{1/2}$ and the proportionality constant increases with $N$ (Fig.~\ref{fig:LLExpc}b).
    
    \item We find intriguing behaviour in the approach to thermalization of  macrovariables, such as density profiles and velocity distributions,  for macroscopic systems of trapped hard rods. For both trap shapes we study thermalization starting from four different types of initial condition, each of which is determined by choosing either a spatially uniform or bimodal (Newton's-cradle-like) position distribution, and choosing either a uniform or a Maxwellian velocity distribution. For each of these four initial conditions, we find that quadratically trapped rods approach different stationary states at large times, none of which corresponds to the conventional Gibbs state (Fig.~\ref{fig:harmNonGibbs}). On the other hand, we find that quartically trapped rods thermalize, eventually reaching the stationary Gibbs state for different initial conditions~(Fig.~\ref{fig:quartNCTherm}).
\end{enumerate}

The paper is organized as follows. 
In Sec.~\ref{models} we describe the model in detail and define the diagnostics that we will be using to characterize its dynamics. In Sec.~\ref{sec:num}, we discuss the numerical methods employed. In Sec.~\ref{sec:results}, we
present the results of extensive molecular dynamics simulations of trapped hard rods. We conclude and discuss some open questions in Sec.~\ref{sec:conc}.

\section{Models and definitions}
\label{models}
We consider one-dimensional hard rods of length $a$ and unit mass in a confining potential, given by the Hamiltonian
\begin{equation}
\label{H}
 H(\{x_i,v_i\}) = \sum_{i=1}^N \left[ \frac{v_i^2}{2} + U(x_i) \right] + \sum_{i=1}^{N-1} V(|x_{i+1} - x_i|),
\end{equation}
where $\{x_i, v_i\}$ denote the position and the momentum of the $i^{th}$ rod  such that $x_{i+1} \ge x_i +a$ for $1 \leq i \leq N-1$. We consider a confining potential of the form 
\begin{equation}
    U(x)=k \frac{x^\delta}{\delta}
\end{equation}
with two values of $\delta$
\begin{equation}
\delta = \begin{cases}2\quad \text{for a quadratic trap}\\ 4 \quad \text{for a quartic trap}
\end{cases}
\end{equation}
The  interaction term  for hard rods is of the form
\begin{equation}\label{hrd:inter}
V(r) = 
\begin{cases}
0 & \text{~for~} \quad r > a \\
\infty & \text{~for~} \quad r \leq a.
\end{cases}
\end{equation}

Under the resulting Hamiltonian dynamics the rods collide elastically with their neighbours, upon which they exchange momenta instantaneously. In between collisions, the rods move independently in the trap potential.  Scaling distances and time by the natural length and time scales,  $a$ and  $\tau=1/\sqrt{k a^{\delta-2}}$, respectively,  one finds the  total energy of the system is given by 
\begin{equation}
\label{E}
E = k a^{\delta} \sum_{i=1}^N \left[ \frac{\dot x_i^2}{2} + \frac{x_i^\delta}{\delta} \right].
\end{equation}
The minimum energy, $E_{\rm m},$ of the system is attained by a close-packed configuration centred at the origin, with all particles at rest. It is clear that $E_m \sim k a^\delta N^{\delta +1}$.
We are interested in observing thermalization at high enough temperatures such that the central density of the gas is reduced from this close-packed density by a factor of order one or more. This requires excitation energy $E_{\rm ex}=E-E_{\rm m}$ of the same order as $E_{\rm m}$ or larger. 
From Eq.~\eqref{E}, we see that the only relevant parameters in the system are the rescaled energy~\cite{kethepalli2022finite}
\begin{equation}
{\rm e} = \frac{E}{{N^{\delta+1} k\, a^\delta}}
\label{eq:cdef}
\end{equation}
and $N$. 
In the following, without loss of generality, we can set $a=1, k=1$ and  compute various physical quantities for different values of the parameters ${\rm e}$ and $N$.  We further note that for the quadratic case, there is a second conserved quantity 
\begin{equation}
E_{\rm cm} = \frac{1}{2N^2} \left[\left(\sum_i x_i\right)^2 + \left(\sum_i v_i\right)^2 \right],
\label{eq:ecm}
\end{equation}
beyond the total energy, which is the energy of the centre of mass~\cite{cao2018incomplete}.  The centre of mass moves autonomously, and the relative motion of the rods is independent of that of the centre of mass, so without loss of generality for the quadratic trap we can restrict to  $E_{\rm cm}=0$. Note that this also implies that $X_{\rm cm}=\sum_i x_i=0$ and $P_{\rm cm}=\sum_i v_i=0$ are separately conserved.

For these systems, we compute the finite time Lyapunov exponent, $\lambda(t)$, and its infinite time limit, $\lambda$, defined respectively as
\begin{align}
\label{Lyapunov}
\begin{split}
 \lambda(t) &=  \lim_{\epsilon \to 0} \frac{1}{t} \ln \left|\frac{d_t}{\epsilon} \right|, \\
 \lambda &= \lim_{t \to \infty} \lambda(t),
 \end{split}
\end{align}
where $d_0 = \epsilon$ is the  separation between the two initial phase-space points, and $d_t$ is their separation at time $t$. For chaotic systems $\lambda > 0$, which represents the exponential divergence of phase-space trajectories for an infinitesimally small initial separation. 
 In fact, it is possible to write a linearised dynamics for the variable $z_t=d_t/\epsilon$ in the $\epsilon \to 0$ limit, which provides an accurate method for computing $\lambda$. We use this method for computing Lyapunov in the quadratic case, whereas for the quartic case we compute it directly  from the evolution of  two different initial conditions. In both cases we use the widely used numerically efficient method due to Benettin, Galgani and Strelcyn~\cite{PhysRevA.14.2338}. To probe thermalization, we compute the (running) time average of the scaled kinetic temperature of the individual hard rods defined as 
\begin{equation}
c_i=\frac{T_i}{N^{\delta}},\quad \text{where}\quad T_i(t) = \frac{1}{t} \int_0^t dt' v_i^2(t'),
\label{eq:scaled_t}
\end{equation}
and check for equipartition.

To study the relaxation dynamics and equilibration to a Gibbs state,  we  compute the spatial density profile $\rho(x,t)$ and the velocity distribution $P(v,t)$ defined as:
\begin{eqnarray}
\label{rhox}
\rho(x,t) &=& \sum_{i=1}^N \left< \delta(x-x_i(t))\right>,\\
P(v,t) &=& \sum_{i=1}^N  \left<\delta(v-v_i(t))\right>.
\label{Pv}
\end{eqnarray}
where $<\cdots>$ denotes an average over  many initial microscopic states with the same initial density profiles and velocity distributions, drawn from a microcanonical ensemble with constant energy ${\rm e}$ and $E_{\rm cm} = 0$. Details of the preparation of these initial states are given below in Sec.~\ref{sec:num}. If the system thermalizes to a Gibbs state, then one expects that $\rho(x)$ will be the same as the equilibrium distribution obtained from 
Monte-Carlo simulations whose temperature is fixed so that the average energy (appropriately scaled) equals ${\rm e}$. The corresponding velocity distribution $P(v)$ will be Gaussian at the same temperature.

\section{Numerical methods}
\label{sec:num}
In this section, we outline the various numerical methods and conventions that we will use both in and out of equilibrium.

\textit{Time evolution:} For the quadratic case ($\delta=2$), one can evolve the equations of motion using exact and numerically efficient event-driven molecular dynamics (EDMD).  For the quartic trap ($\delta=4$) case, we employ standard molecular dynamics (MD) simulations using a symplectic velocity-Verlet integration scheme. 
 During collision events, we exchange the velocities of the particles at the first instant that any two adjacent rods overlap, defined as $x_{i+1}-x_i<a$. To ensure the accuracy of this approximation, we use a very small time increment $dt=10^{-6}$. 

\textit{Stochastic momentum exchange dynamics (SMED): }To sample initial conditions uniformly over the phase space from a microcanonical ensemble with ${\rm e}$ fixed and $E_{\rm cm}=0$, we allow momentum exchange of randomly chosen pairs of neighbouring particles at random times in addition to the usual Hamiltonian dynamics. This stochastic process conserves the total momentum and energy of the system. For the quadratic trap case, this stochastic momentum exchange dynamics (SMED) also conserves the centre of mass energy $E_{\rm cm}$. The SMED exhibits the expected equipartition of energy (flat temperature profiles) and insensitivity to initial conditions, both of which are consistent with ergodicity.

\textit{Initial state preparation:} 
To check the initial condition dependence of the maximal Lyapunov exponent $\lambda$ and its distribution we used microcanonical initial conditions generated by the SMED.

To check thermalization, we prepare the system with specified nonequilibrium spatial density profiles $\rho(x)$ and velocity distributions $P(v)$ consistent with given values of ${\rm e}$ and $E_{\rm cm}=0$. This is achieved via the following protocol. First, we distribute the rods spatially in accordance with the required density profile $\rho(x)$, imposing the hard-rod constraint and fixing the centre of mass at $x=0$. We then compute the total potential energy $E_{\rm p}$ for this configuration and subtract it from the total energy $E$ to obtain the total kinetic energy $E_{\rm k}$. The velocities are drawn from the distribution $P(v)$, and then shifted and rescaled by appropriate factors so that the centre of mass velocity vanishes and the total kinetic energy is exactly $E_{\rm k}$.
In this work we consider two non-thermal choices of $\rho(x)$: either uniform over a finite width (denoted U), or a Newton's-cradle-like profile consisting of two uniform blobs, each of finite width and separated by an $O(N)$ distance (denoted Nc). For the velocities, we consider two choices of $P(v)$: either uniform (denoted U) or Maxwellian (denoted Mx). This leads to four possible choices of non-equilibrium initial conditions: (i) U-U, (ii) U-Mx, (iii) Nc-U, (iv) Nc-Mx.

\begin{figure}[t]
	\centering
        \includegraphics[scale=0.5]{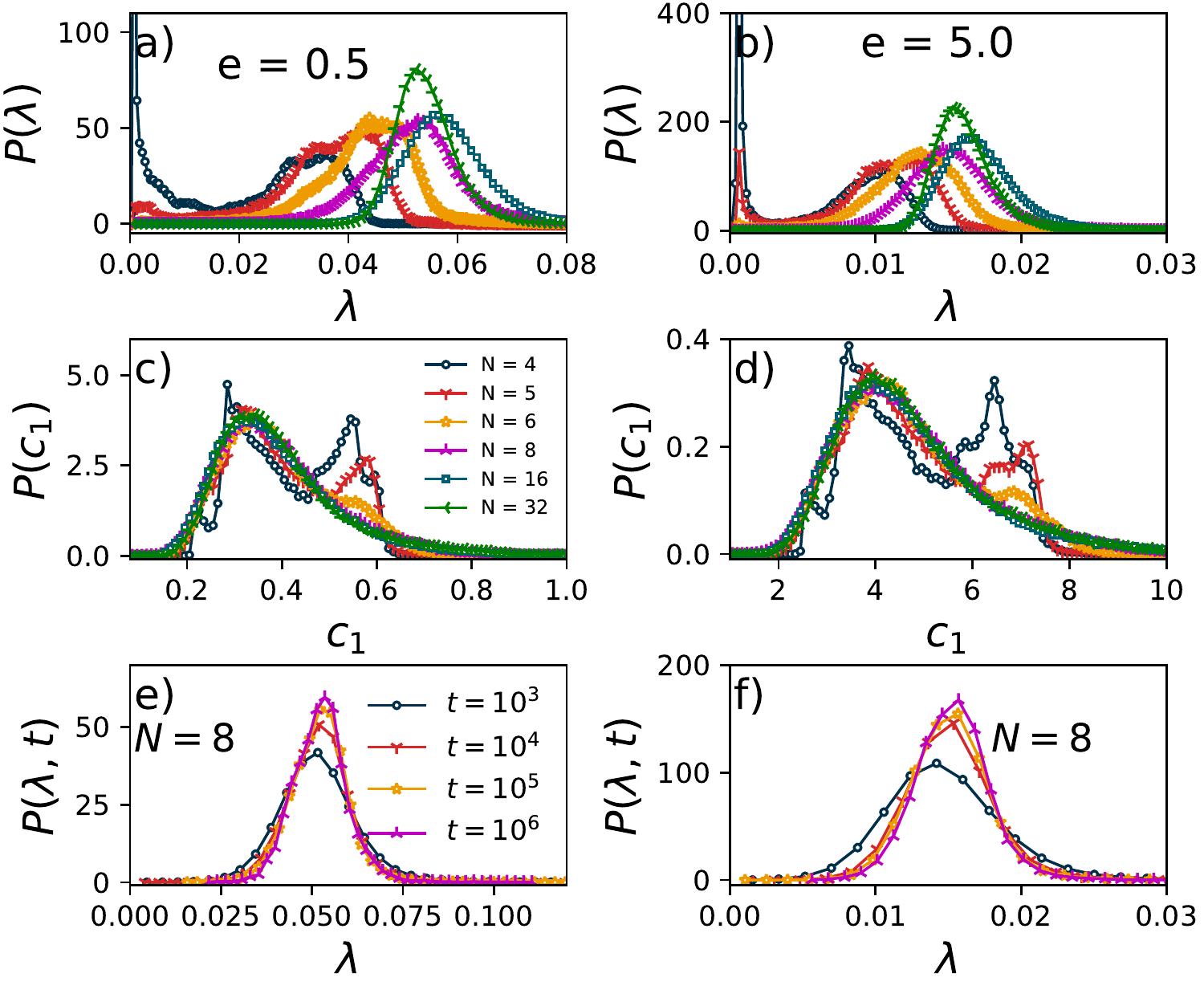}
 \caption{(a,b,c,d) Distribution of the maximal Lyapunov exponent $\lambda$ and rescaled temperature $c_1$ of the left-most rod, for $4 \leq N \leq 32$, computed at time $t = 10^5$. The system sizes corresponding to different plots are provided in sub-plot (c). We find a significant breakdown of ergodicity for hard rods in this quadratic trap. (e,f) Distribution of $\lambda$ for $N = 8$ at different times, $10^3 \leq t \leq 10^6$, which shows that $P(\lambda,t)$ approaches a steady limiting distribution at late times. The initial conditions for all the plots are generated using the SMED: (a,c,e) for ${\rm e} = 0.5$ and (b,d,f) for ${\rm e} = 5.0$. }
	\label{fig:harmproblambda}
\end{figure}

\section{Results on Chaos, ergodicity and thermalization}
\label{sec:results}

 As mentioned earlier, one na{\"i}vely expects  that  the presence of the trap makes the system chaotic ($\lambda > 0$), ergodic (no long-time dependence on the details of the initial condition), and non-integrable (strictly fewer than $N$ independent integrals of motion). In the following we investigate these properties in detail by computing the Lyapunov exponent and kinetic temperatures for different $N$ in quadratic ($\delta=2$) and quartic ($\delta=4$) trapping potentials.

\subsection{Chaos and ergodicity}
It is easy to see that the dynamics of hard rods with $N=2$ is integrable for the  quadratic trap because of the presence of the second conserved quantity $E_{\rm cm}$. This is however not the case  in a quartic trap,  as will be elaborated below.

\noindent
\paragraph*{$N=3$ rods (quadratic trap):} We first consider the case  of $N = 3$ rods in the quadratic trap with $E_{\rm cm}=0$. We find that the systems displays features akin to integrable systems as exhibited by the existence of non-chaotic trajectories with Lyapunov exponents decaying as $\lambda(t) \sim \log (t) /t $ (Fig.~\ref{fig:LLExpN3}a). This is similar to  integrable models such as the Toda chain~\cite{ganapa2020thermalization}. The Poincar\'e sections are shown in Fig.~\ref{fig:LLExpN3}b where we observe regular patterns consistent with Fig.~\ref{fig:LLExpN3}a. 

\noindent
\paragraph*{$N=2$ rods (quartic trap):} In striking contrast to the above case, the behaviour of even $N=2$ rods in a quartic trap shows both chaotic and regular trajectories, as depicted in Fig.~\ref{fig:LLExpN3}c. This observation is consistent with the Poincar\'e sections shown in Fig.~\ref{fig:LLExpN3}d, where we observe that the phase space of two hard rods can have disjoint chaotic regions (scattered) and non-chaotic (regular) islands. However, our observations indicate that the  phase space volume of the regular island is much smaller than that of the chaotic region even for $N=2$.

\begin{figure}[t]
	\centering
        \includegraphics[scale=0.5]{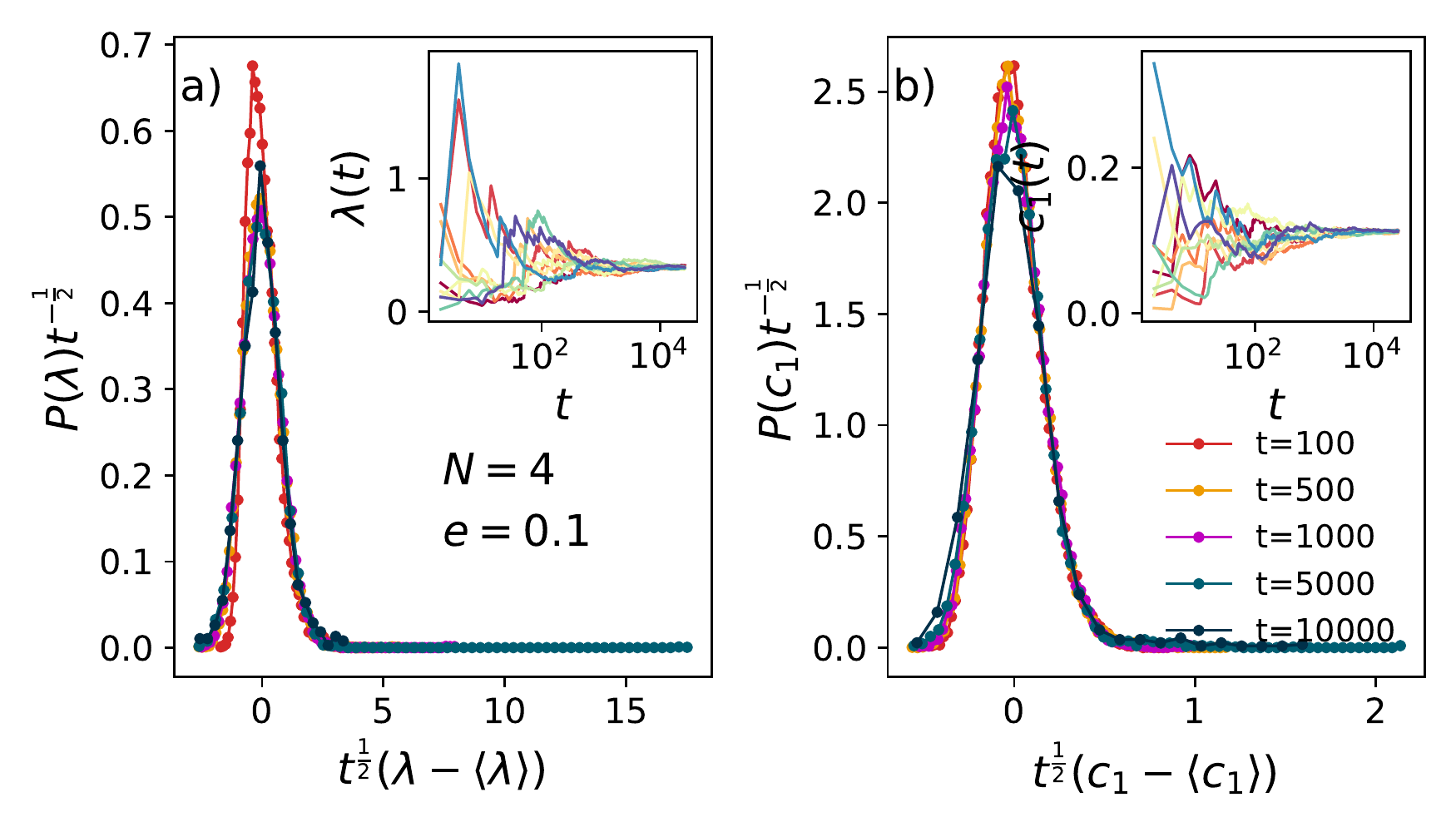}
\caption{Probability distribution of (a) maximum Lyapunov exponents $\lambda$  and (b) $c_1$  for $N=4$ rods in a quartic trap. We observe that both distributions collapse at different times after shifting by their respective means and scaling  by $\sqrt{t}$. This indicates that the width of these two distributions decreases with time and becomes increasingly sharp, in contrast to our findings for the quadratic trap depicted in Fig.~\ref{fig:harmproblambda}. In the insets, we show the time evolution of $\lambda$ and $c_1$ for $20$ different realizations (each colour denotes one initial condition), all of which converge to a unique value at late times, regardless of the initial conditions.}
	\label{fig:quartproblambda}
\end{figure}

\noindent
\paragraph*{$N\geq 4$ rods (quadratic trap):} We find that many trajectories for $N=4$ rods in a quadratic trap are chaotic, although still non-thermalizing. We compute $\lambda(t)$ and  $c_1(t)$ for different  initial conditions (IC) obtained from SMED simulations (see Section~\ref{sec:num}) for two values of the rescaled energy, ${\rm e}=0.5$  and ${\rm e}=5.0$. The results are shown in Figs. ~\ref{fig:N4timeseries}a,c and Figs.~\ref{fig:N4timeseries}b,d, respectively. We find that the values of $\lambda(t)$ and $c_1(t)$ at late times are sensitive to the choice of initial condition. Interestingly, we observe that even for $N=4$ there is a fraction of trajectories for which $\lambda(t)$ decays in time for all numerically accessible times, as for the case of $N=3$ rods (see Figs.~\ref{fig:N4timeseries}a,b). To investigate equipartition we plot $c_i(t)$ for $i=1,2,3,4$ for a single initial condition in Fig.~\ref{fig:N4timeseries}e for ${\rm e=0.5}$, and observe that $c_1(t)=c_4(t)$ and $c_2(t)=c_3(t)\neq c_1(t)$ at late times. This is also observed for ${\rm e=5.0}$ in Fig.~\ref{fig:N4timeseries}f. These observations suggest that the $N=4$ system is chaotic but not ergodic for most choices of initial condition.

To quantify and further investigate the IC dependence and  non-ergodicity in systems with different numbers of rods $N$, we compute the probability distributions $P(\lambda)$ and $P(c_1)$ of the late time values of $\lambda(t)$ and $c_1(t)$, obtained from an ensemble of ICs (once again generated using SMED) for $e=0.5$ (Figs.~\ref{fig:harmproblambda}a,c) and $e=5.0 $ (Figs.~\ref{fig:harmproblambda}b,d). Interestingly, for the distribution $P(\lambda)$, we see a peak near $\lambda=0$ for $N=4$ arising from the  non-chaotic  trajectories observed in Figs.~\ref{fig:N4timeseries}a,b. This peak, however,  decreases sharply with increasing $N$. Further, we observe that  the mean of the distribution $P(\lambda)$ behaves non-monotonically with increasing $N$. On the other hand, the width of the distribution seems to decrease with increasing $N$. The fact that the distributions of both $\lambda$ and $c_1$ are still quite broad even at the largest system size $N=32$ studied is strong evidence for a lack of ergodicity in the system. In order to demonstrate that $t=10^5$ is a sufficiently long time for computing the distributions $P(\lambda)$ in Figs.~\ref{fig:N4timeseries}a,b, we, in  Figs.~\ref{fig:harmproblambda}e,f plot  the distribution of $\lambda(t)$ at different times for $N=8$. We observe that these distributions initially display some narrowing, but seem to converge to a limiting form of finite width at long times. This suggests that the system is genuinely non-ergodic and that the identification  of $\lambda$ with $\lambda(t)$ at $t=10^5$ in Fig.~\ref{fig:harmproblambda}a,b is justified.

\begin{figure}[t]
    \includegraphics[scale=0.55]{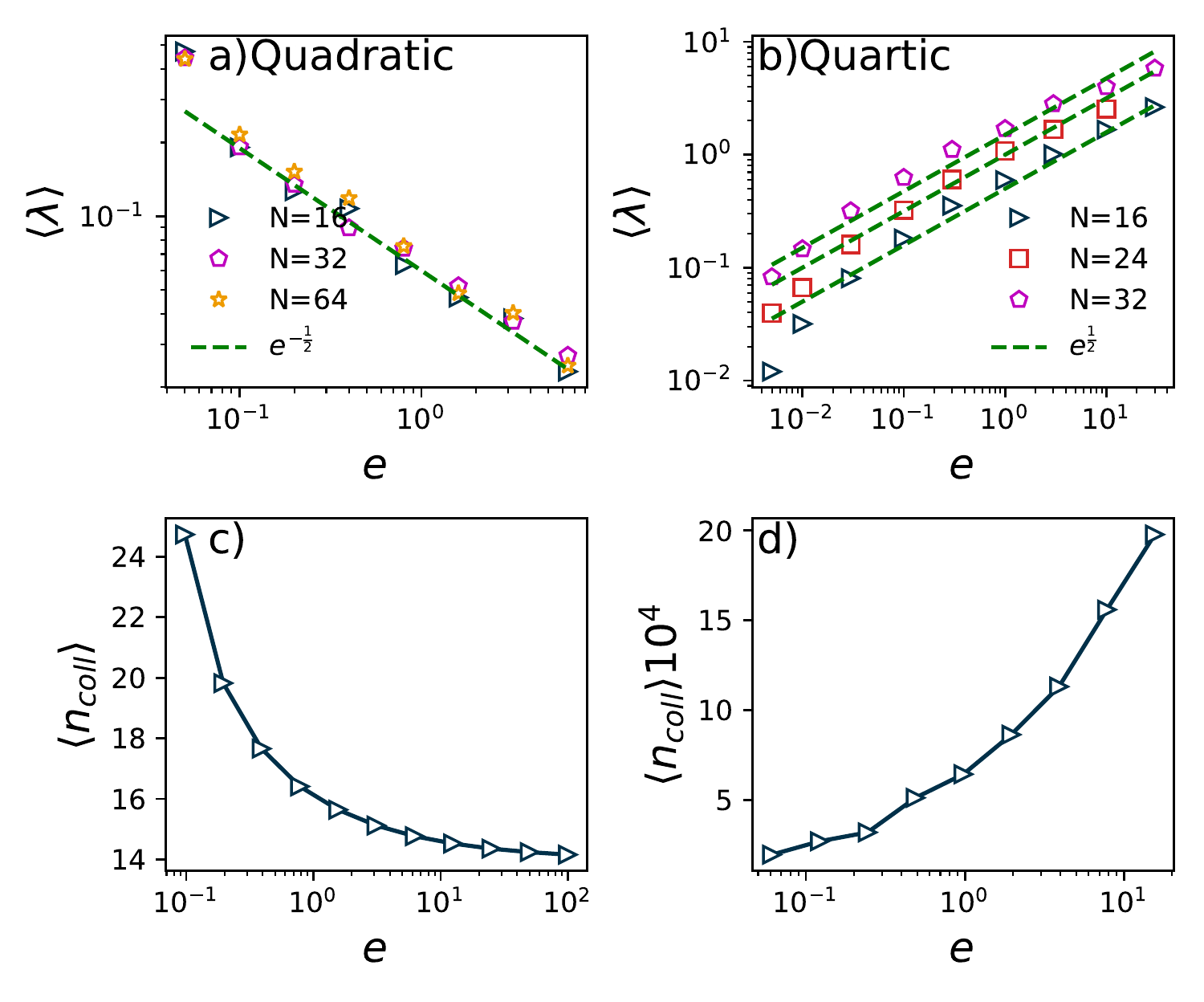}
	\caption{Plot of average maximal Lyapunov exponent ($\langle \lambda \rangle$) with rescaled energy ${\rm e}$ for (a) quadratic and (b) quartic trap. The average number of collisions per unit time $\langle n_{\rm coll} \rangle$ as a function of total energy ${\rm e}$ for $N=8$ for (c) quadratic trap and (d) quartic trap. }
	\label{fig:LLExpc}
\end{figure}

\begin{figure*}[t]
{\includegraphics[width=8.7cm]{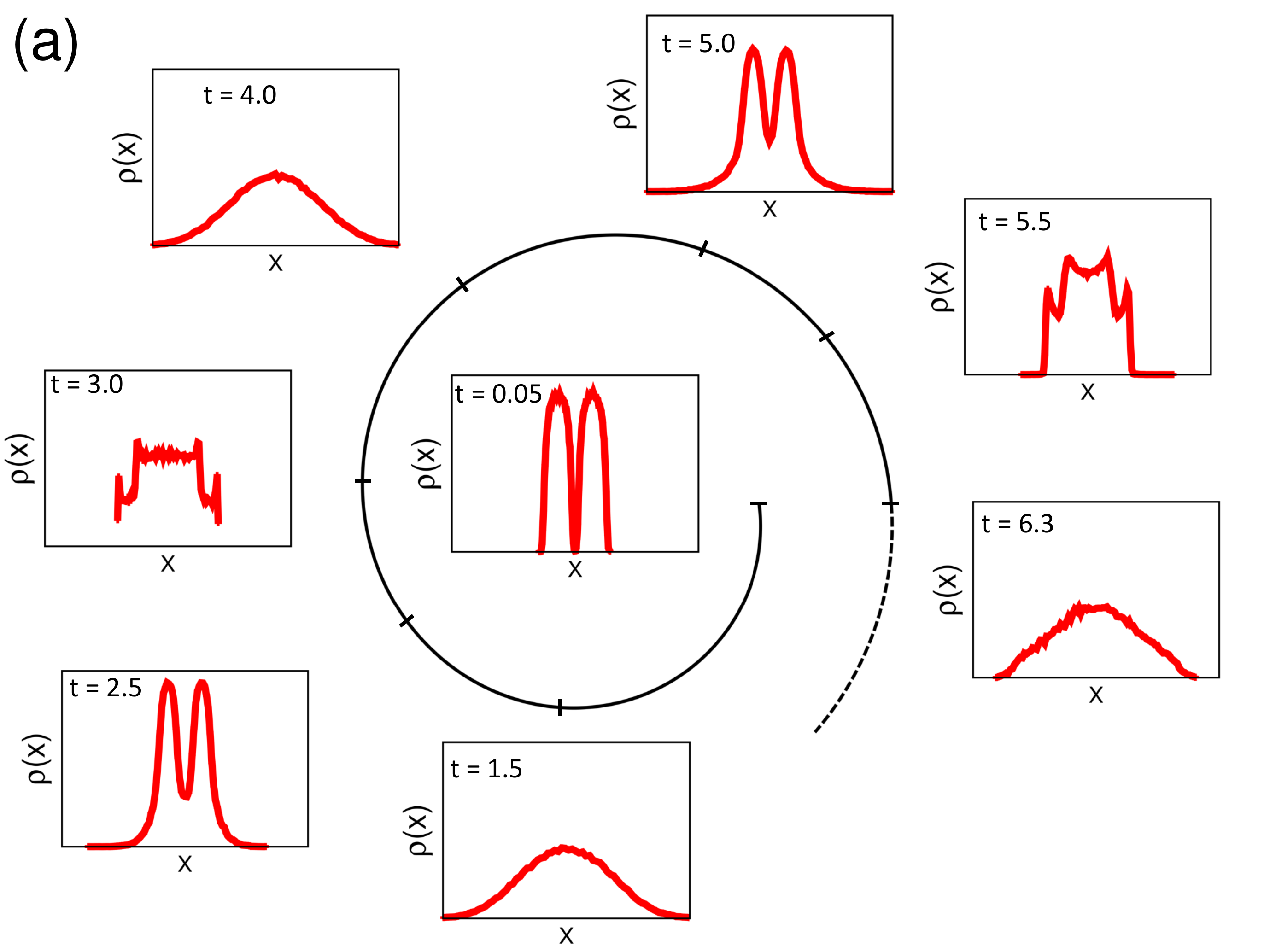}}\hskip0.4cm
{\includegraphics[width=8.7cm]{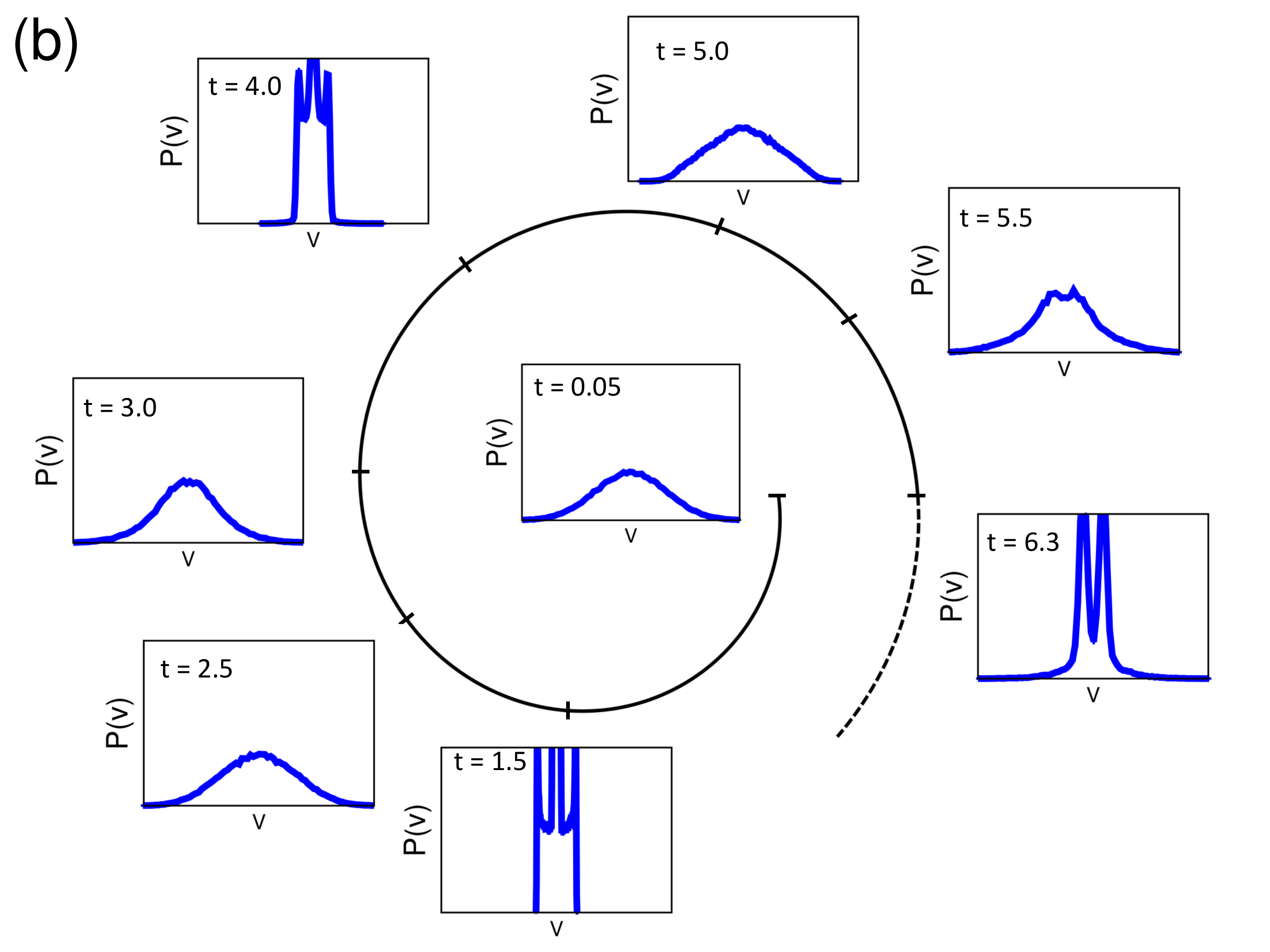}}
	\caption{(a) Density profiles $\rho(x)$ and (b) velocity distributions $P(v)$ in the quadratic trap, with ${\rm e}=0.5$ and $N = 128$, that appear at different times within one time-cycle of the trap, $0 < t \lesssim \tau = 2\pi$. These profiles are obtained starting at $t=0$ from a two-blob density profile and Gaussian velocity distribution ({\it i.e.} Nc-Mx)  of the hard-rods. For all the figures in (a) and (b), the abscissa runs from $-400$ to $400$. For (a) and (b), the ordinate scale ranges are $0-0.007$ and $0-0.01$ respectively. As can be clearly observed, the hard-rod system in a quadratic trap has an initial `breathing-mode' dynamics and exhibits oscillations in the distributions, somewhat resembling a Newton's cradle. For our parameters these oscillations damp out in $O(20)$ cycles. 
 }
	\label{fig:SmallTime}
\end{figure*}

\begin{figure}[h]
        {\includegraphics[width=8.25cm]{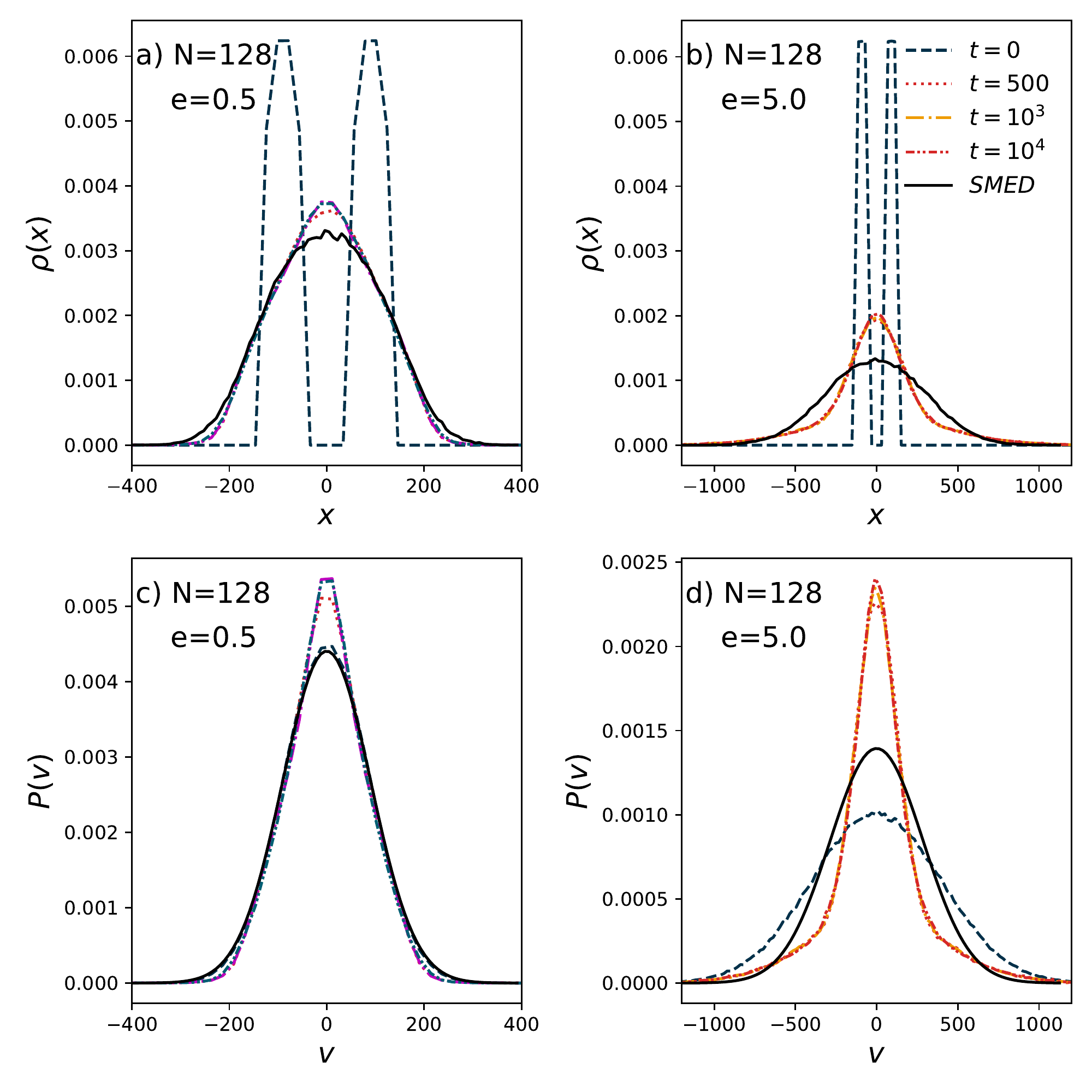}} 
	\caption{Time evolution of density and velocity profiles: (a,c) for ${\rm e} = 0.5$ and (b,d) ${\rm e} = 5.0$ for $N  = 128$ hard rods in a quadratice trap, starting from a Newton's cradle initial condition (i.e. two spatially separated blobs of rods) with Maxwellian velocities (Nc-Mx, in the notation of Sec. \ref{sec:num}). The times simulated are indicated by the legend in (b). These plots illustrate that at late times ($t = 10^4 \tau$) the density profiles and velocity distributions obtained from EDMD converge to forms that differ from those obtained from SMED.}
	\label{fig:harmNCTherm}
\end{figure}

These numerical results are consistent with the following possible scenarios for the quadratic trap: 
\begin{itemize}
      \item The disappearance of the peak in $P(\lambda)$ at $\lambda=0$ with increasing $N$ indicates that any possible KAM-like non-chaotic islands occupy negligible phase-space volume in the limit of large system size.
      
      \item The non-vanishing width of $P(\lambda)$ and $P(c_1)$ for the simulated values of $N$ suggests the existence of multiple chaotic islands with distinct values of $\lambda$ and $c_1$ in a given microcanonical shell.
      
      \item These chaotic islands could arise either from extra conserved quantities or from strong kinetic constraints (e.g. high entropy barriers) that prevent movement between different islands.  In the former case, we expect that the width of the distributions $P(\lambda)$ and $P(c_1)$ will not go to zero even for long times and large $N$. In the latter case, these distributions will eventually become sharp at sufficiently long times, yielding unique values of $\lambda$ and $c_1$ for any $N$. Our numerical results in Fig.~\ref{fig:harmproblambda}(e,f) are in closer agreement with the former scenario.
\end{itemize}

\paragraph*{$N\geq 4$ rods (quartic):}  For the quartic trap, numerically obtained distributions for $P(\lambda(t))$ and $P(c_1(t))$ are shown in Figs.~~\ref{fig:quartproblambda}a and ~\ref{fig:quartproblambda}b respectively, for different times from $t=100$ to $t=10^4$. In contrast to the quadratic trap, we find that both these distributions are sharply peaked, and that their width decreases with time as $\sim{t^{-1/2}}$ (see the scaling in Fig.~\ref{fig:quartproblambda}). This suggests that hard rods in a quartic trap thermalize. This conclusion is supported by the insets of these figures, which demonstrate that $\lambda(t)$ and $c_1(t)$ converge  to unique values (within statistical fluctuations) for different initial conditions. Thus our numerical simulations find negligible dependence of the late-time dynamics on initial conditions, which is evidence for thermalization, and consistent with ergodicity (testing the latter directly would require a more detailed analysis of individual phase-space trajectories).

\begin{figure}[h]
        {\includegraphics[width=8.25cm]{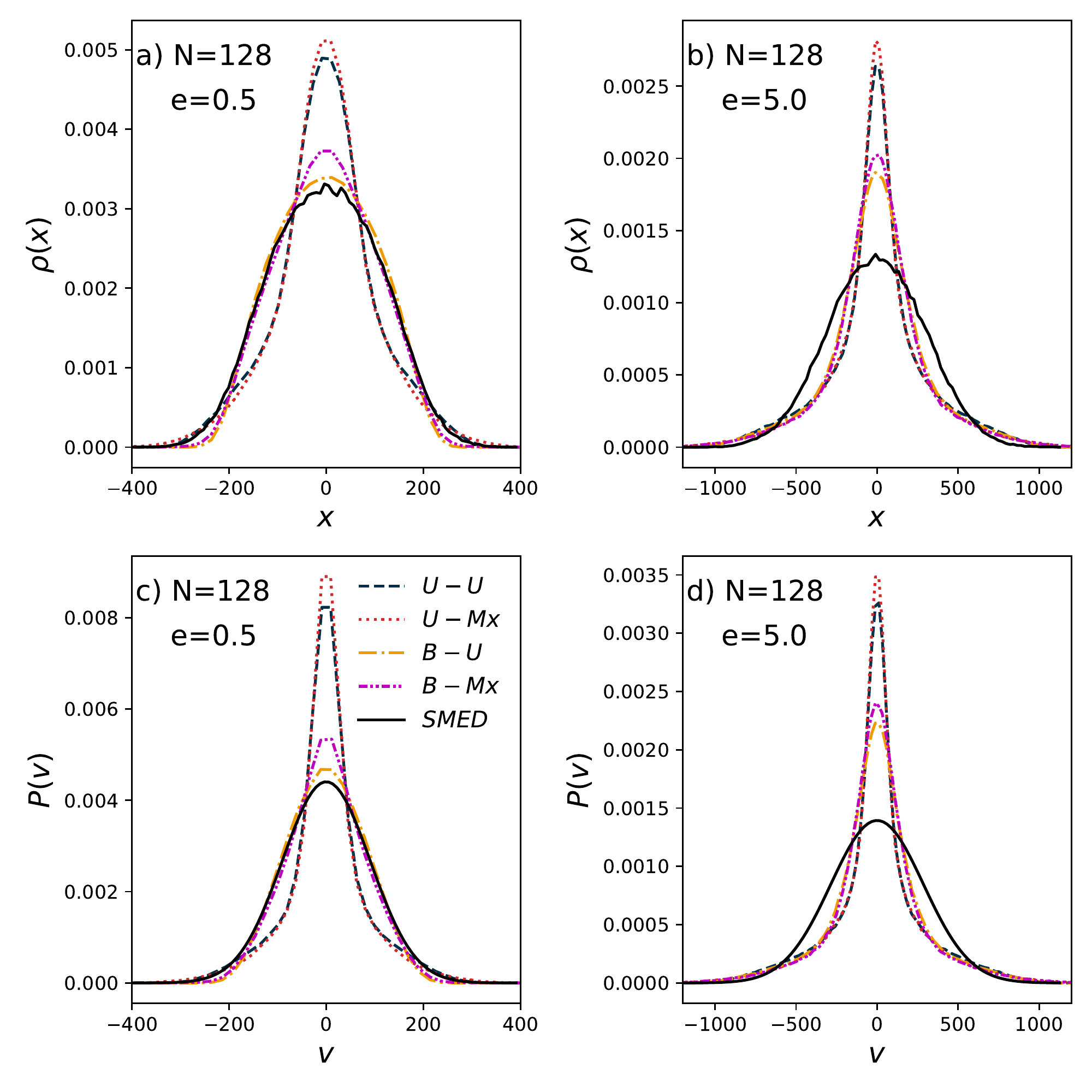}} 
    \caption{In this figure, we investigate the initial condition dependence of the late time ($t = 10^4 \tau$) distributions obtained from EDMD of  hard rods in a quadratic potential. These are compared with thermal predictions obtained from SMED. We compare the density and velocity profiles of $128$ rods at two energies (and : (a-c) ${\rm e} = 0.5$ and (b-d) ${\rm e} = 5.0$. We use four different initial conditions: (i) uniform density and uniform velocity distribution (blue dashed line), (ii) uniform density and Gaussian velocity distribution (red dotted line), (iii) Newton's cradle density and uniform velocity distribution (yellow dashed-dotted line), and (iv) Newton's cradle density and Gaussian velocity distribution (magenta dashed-double-dotted line). We find that neither the density profile nor the velocity distribution agree between EDMD and SMED (black solid line), even at long times $t = 10^4 \tau$. We also observe that late-time density profiles depend on the choice of initial condition for both temperatures.}
	\label{fig:harmNonGibbs}
\end{figure}
\begin{figure}[h]
	\includegraphics[width=8.25cm]{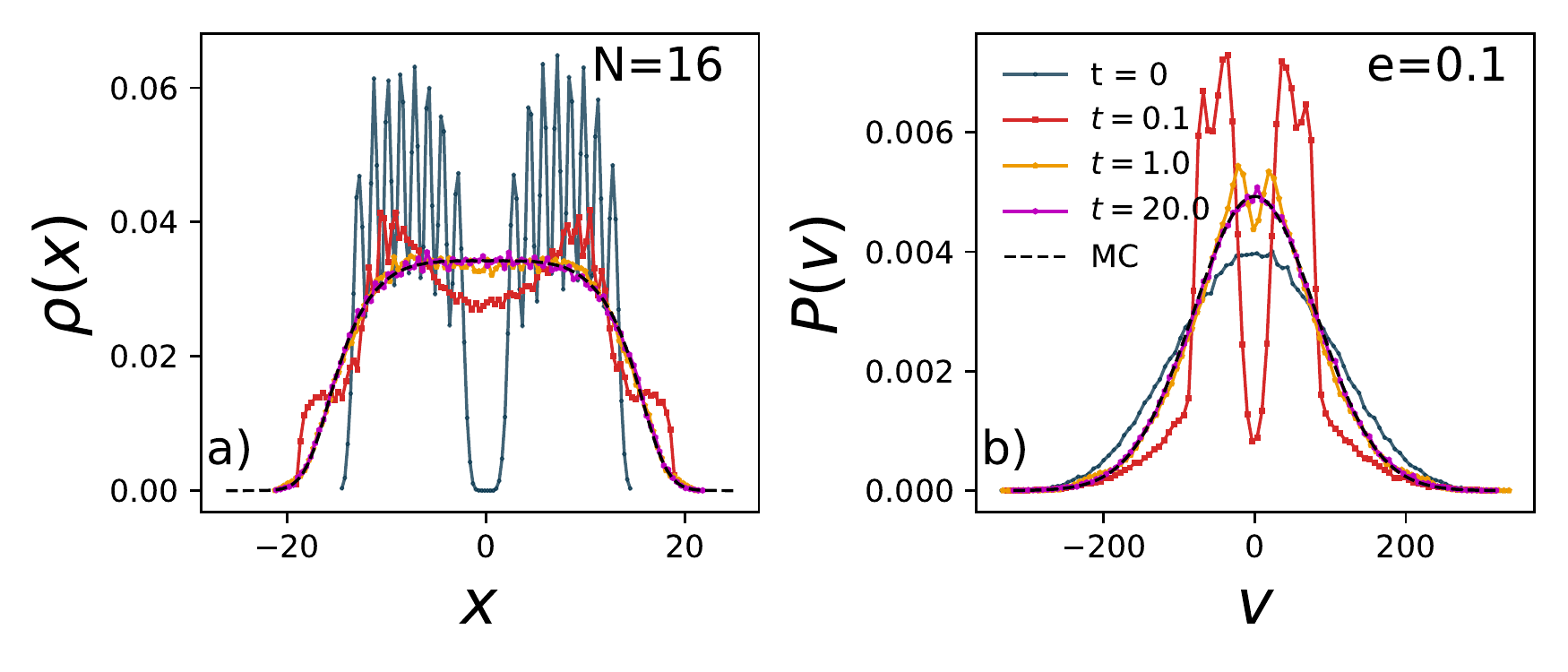}\\
	\includegraphics[width=8.25cm]{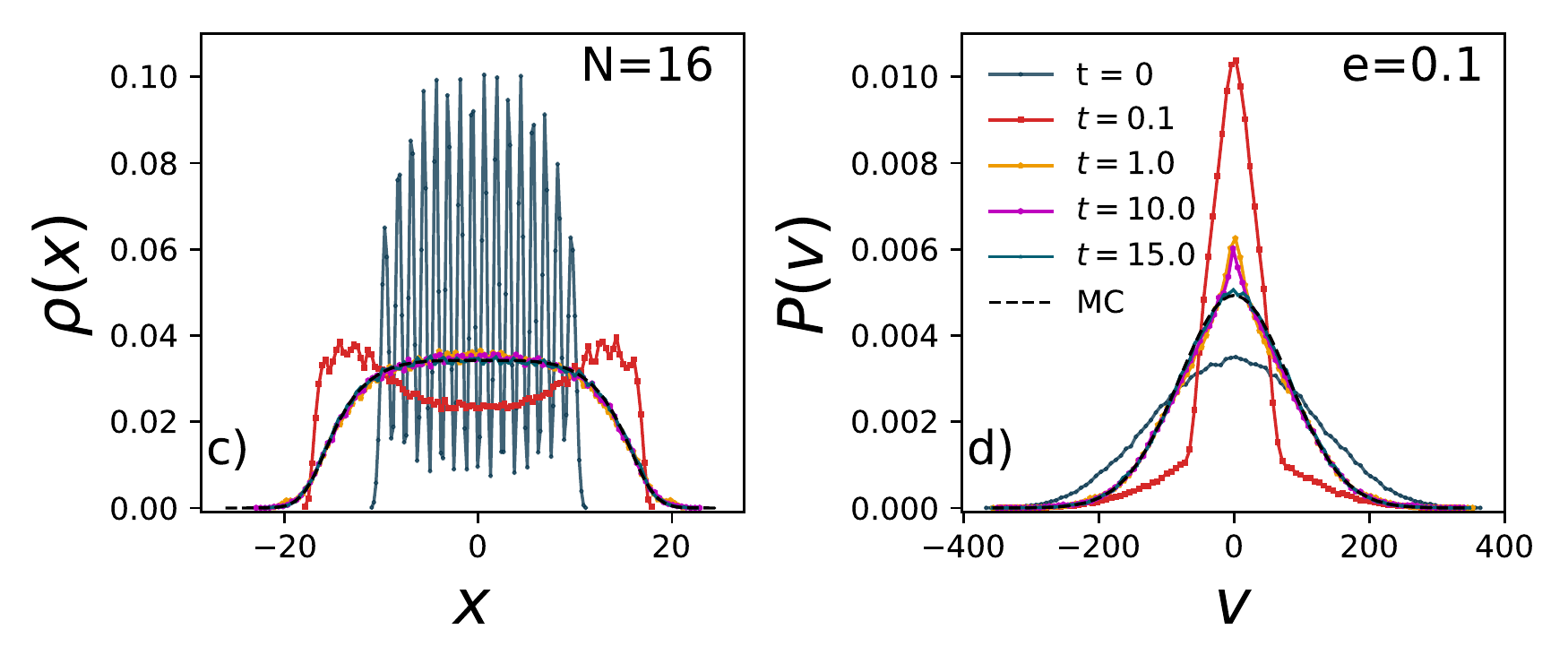}
	\caption{Time evolution of (a-c) MD density and (b-d) velocity distribution for hard-rods in a quartic trap, starting from (a-b) Nc-Mx and (c-d) U-MX initial condition, compared with Monte-Carlo profiles for $N = 16$ and ${\rm e} = 0.10$. In this case, the MD profile converges to the Monte Carlo (MC) result appreciably fast, and the velocity distribution approaches a Gaussian at late times, as expected for a non-integrable system.}
	\label{fig:quartNCTherm}
\end{figure}

\subsection{Energy dependence of chaos}
In this section, we investigate how the mean maximal Lyapunov exponent $\langle \lambda \rangle$ (obtained from the distributions in Figs.~\ref{fig:harmproblambda}a and~\ref{fig:harmproblambda}b) depends on the rescaled energy ${\rm e}$ and $N$ for both traps.  We observe that in the case of the quadratic trap, $\langle \lambda \rangle$ roughly saturates to a non-zero value at large $N$ for a fixed value of ${\rm e}$. In  Fig.~\ref{fig:LLExpc}a we plot these saturation values as a function of ${\rm e}$ where one observes that $\langle \lambda \rangle$ decreases with ${\rm e}$ as $\sim 1/\sqrt{{\rm e}}$ at large ${\rm e}$. A similar decrease of $\langle \lambda \rangle$ with increasing energy has been reported earlier for soft rods in a quadratic trap \cite{dong2018classical}.  For the quartic trap, in contrast to the quadratic case, $\langle \lambda \rangle$ does not appear to converge with increasing $N$ for the range of $N$ values studied here. For fixed $N$, $\langle \lambda \rangle$ grows with increasing ${\rm e}$ as $\sim\sqrt{{\rm e}}$ for large ${\rm e}$ as can be seen from Fig.~\ref{fig:LLExpc}b. This square-root dependence of $\lambda$ on temperature is also observed in other non-integrable systems~\cite{murugan2021many,kurchan2018quantum}.

To understand this intriguing dependence of $\lambda$ on ${\rm e}$ better, we compute the average number of collisions per unit time $\langle n_{coll} \rangle$ in both traps, for a fixed $N = 8$ and for different values of the  energy ${\rm e}$. These are shown in Figs.~\ref{fig:LLExpc}c and~\ref{fig:LLExpc}d for the quadratic and the quartic trap respectively. From Fig.~\ref{fig:LLExpc}c we find that $\langle n_{coll} \rangle$ decreases in  the quadratic trap as ${\rm e}$ is increased. Thus, as the energy is increased the hard rod gas expands and collisions become rarer. We expect that this reduced rate of collisions is responsible for the decrease in $\langle \lambda \rangle$ with increasing ${\rm e}$ for the quadratic trap. In contrast, we find for the quartic trap that $\langle n_{coll} \rangle$ increases as ${\rm e}$ is increased (see Fig.~\ref{fig:LLExpc}d), which may cause the increase of $\langle \lambda \rangle$ with ${\rm e}$.

\subsection{Thermalization in macroscopic systems}
In previous sections, we studied the chaos and ergodicity properties of hard rods in quadratic and quartic traps. For quadratic traps, we found numerical evidence that for large $N$ the system is chaotic but not ergodic, while for quartic traps we found that the system was both chaotic and thermalizing (and most likely ergodic). A notable feature of the quadratic trap is that the dynamics becomes \textit{less} chaotic as the rescaled energy is increased. 

Whether these results have any bearing on thermalization in macroscopic systems is a nontrivial question, which we now address. We will study this question by looking at the time evolution of non-equilibrium density profiles and velocity distributions of trapped hard rods (evolving under Hamiltonian dynamics) and checking whether these relax to the Gibbs state.

To this end, we compute $\rho(x,t)$ and $P(v,t)$, as defined in Eqs.~\eqref{rhox} and \eqref{Pv}, as a function of time for four choices of initial condition (see Sec.~\ref{sec:num}) with fixed values of ${\rm e}$ and $E_{\rm cm}=0$. In Figs.~\ref{fig:SmallTime}a and b we show $\rho(x)$ and $P(v)$ for small times $0 < t \lesssim \tau = 2\pi$, with $N = 128$ hard rods in the quadratic trap, starting from IC Nc-M, {\it i.e}, from a Newton's cradle initial condition in space (two spatially separated blobs of rods) with velocities chosen from a Maxwell distribution. It is clear that the rods, starting from a two-blob initial condition (at $t = 0$), go through ``breathing'' dynamics and exhibit large oscillations in their density profiles and the velocity distributions. As the system ``breathes'', the density profile goes through different intriguing shapes that are shown in Fig.~\ref{fig:SmallTime}a.  Such transients in the finite-time dynamics of trapped integrable systems are well documented by now~\cite{kinoshita2006quantum,cao2018incomplete,caux2019hydrodynamics,bastianello2021hydrodynamics}.

After these initial transients, the position and velocity distributions begin to approach a stationary state. We plot the single-particle distributions for $N=128$ hard rods in a quadratic trap at late times $t=500, \, 10^3, \,10^4$.  These distributions are shown in Figs.~\ref{fig:harmNCTherm}a-d for ${\rm e}= 0.5$ and ${\rm e} = 5.0$.  To check  whether or not the rods thermalize in the long-time limit, we also plot the corresponding single-particle distributions obtained from SMED, which are expected to recover the microcanonical ensemble.

Strikingly, in the quadratic trap, we find that the density profile obtained from the microscopic dynamics even at the longest accessible times, $t = 10^4\tau$, (where $\tau = \frac{2 \pi}{\omega}$ is the time period of the trap) is very different from the SMED prediction. The velocity distribution is also found to differ from the SMED prediction, for both ${\rm e} = 0.5$ and ${\rm e} = 5.0$. Thus the hard rod gas does not thermalize in quadratic trap even at the very longest accessible times. This is consistent with earlier work, which found that the long-time steady state of quadratically trapped hard rods was a non-thermal stationary solution to ballistic-scale GHD on comparable timescales~\cite{cao2018incomplete}. It appears that for smaller ${\rm e}$, the density and velocity profiles are closer to the equilibrium forms obtained from SMED.  Thus, quite intriguingly, we find that quadratically trapped hard rods at a higher rescaled energy ${\rm e}$ are less chaotic, retain the memory of their initial conditions for longer, and show greater reluctance to thermalize than systems at lower ${\rm e}$.

To argue convincingly against thermalization, we must further check that the late-time behaviour of the system is sensitive to the choice of initial condition. In Fig.~\ref{fig:harmNonGibbs}, we investigate the late-time behaviour of hard rods in a quadratic potential for several initial conditions and compare them with the corresponding thermal predictions from SMED. The four different initial conditions (see Sec.~\ref{sec:num}) considered are (i) uniform density and uniform velocity distribution (U-U), (ii) uniform density and Maxwell velocity distribution (U-Mx), (iii) Newton's cradle density and uniform velocity distribution (Nc-U), and (iv) Newton's cradle density and Maxwell velocity distribution (Nc-Mx). We find that the neither the density profiles nor the velocity distributions of the late-time microscopic dynamics are consistent with SMED. Remarkably, even the late-time distributions obtained by evolving different initial conditions under the microscopic dynamics are distinct from one another, implying non-ergodicity.

In sharp contrast, hard rods in a quartic trap thermalize rapidly to a Gibbs state, regardless of the choice of initial condition. This is shown for two macroscopically distinct initial conditions in Figs.~\ref{fig:quartNCTherm}a and b (for the NC-Mx initial condition) and Figs.~\ref{fig:quartNCTherm}c and d (for the U-Mx initial condition), where long-time density and velocity distributions obtained from the microscopic dynamics are compared with the expected equilibrium distributions. We observe excellent agreement for both choices of initial condition.

\begin{figure}[t]
	{\includegraphics[width=8.27cm]{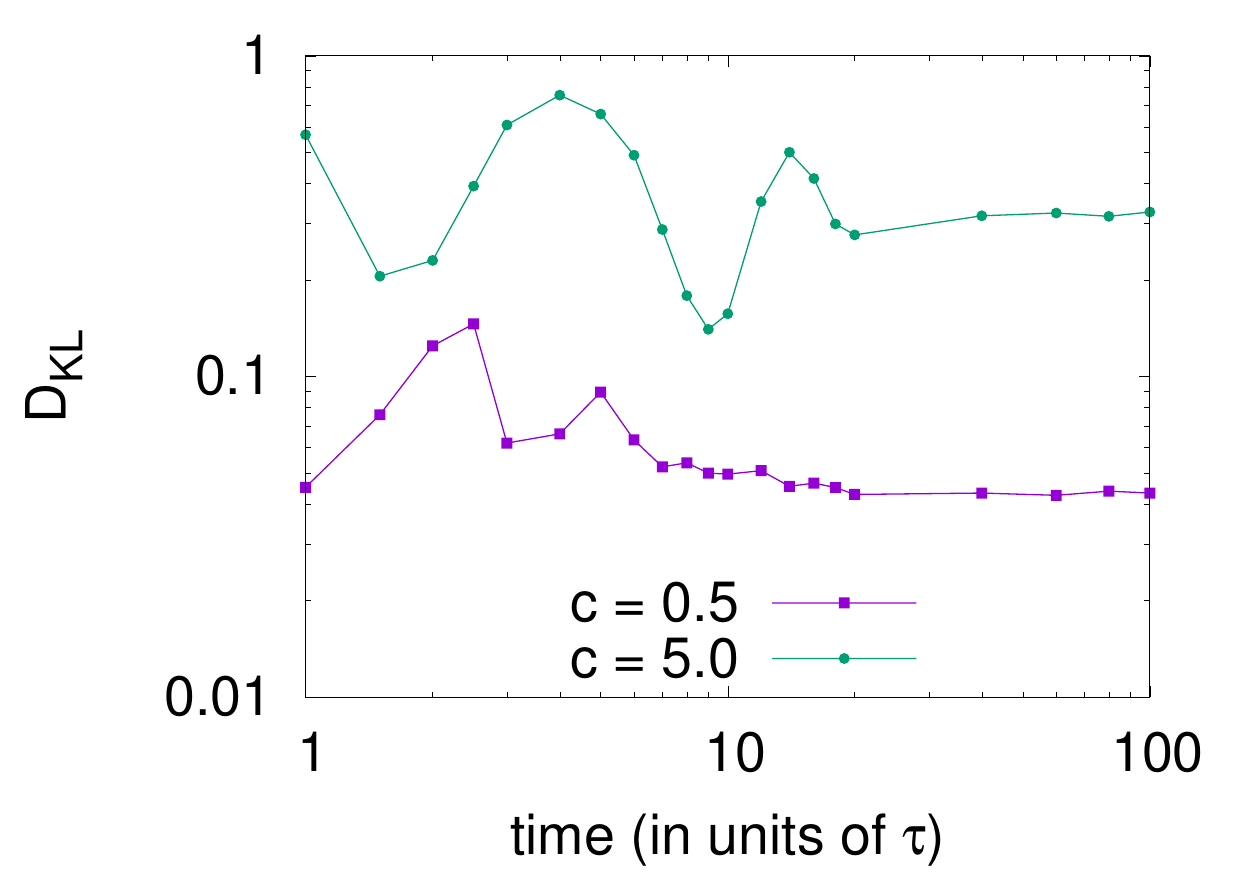}}
	\caption{Time evolution of the distance measure, $D_{KL}$, between $\rho(x)$ and $\rho_{\rm SMED}(x)$ for ${\rm e} = 0.5$ and ${\rm e} = 5.0$. For both EDMD and SMED the system initially is prepared in Nc-Mx initial condition. The oscillation at small times are consistent with oscillation observed in Fig.~\ref{fig:SmallTime}. The saturation of $D_{KL}$ to non-zero values at large times indicate lack of thermalization to Gibbs state.}
	\label{fig:KL}
\end{figure}

To characterize the lack of thermalization of the hard-rods in a quadratic trap in a more quantitative manner, we characterize the `distance' of the EDMD density profiles $\rho(x)$, from the expected equilibrium distributions $\rho_{SMED}(x)$ (obtained from SMED), using the (symmetrized) Kullback-Liebler divergence, defined as
\begin{equation}
D_{KL}(\rho, \rho_{_{SMED}}) = \sum_x \frac 12\left[\rho \log \left|\frac{\rho}{\rho_{_{SMED}}} \right| + \rho_{_{SMED}} \log \left|\frac{\rho_{_{SMED}}}{\rho} \right|\right].
\end{equation}
The Kullback-Liebler divergence as a function of time, for two different ${\rm e}$ values, is shown in Fig.~\ref{fig:KL}. As anticipated, $D_{KL}$ for ${\rm e} = 5.0$ is clearly larger than $D_{KL}$ for ${\rm e} = 0.5$. Furthermore, $D_{KL}(t)$ at long times ($t \sim 10^4\tau$) seems to saturate to a non-zero value, implying a lack of thermalization.

\section{Conclusions and Outlook}
\label{sec:conc}
In this paper, we have investigated chaos, ergodicity and thermalization for one-dimensional gases of classical hard rods in quadratic and quartic traps. Our work demonstrates that thermalization properties are radically different between quadratic traps and quartic traps. In the quadratic case, even though the system has a positive Lyapunov exponent confirming that integrability is broken, the dynamics nevertheless appears to be non-ergodic and fails to thermalize on the accessible timescale.  This is markedly different from expectations for conventional non-integrable classical many-body systems. Our main findings for the case of $N>3$ hard rods are summarised in Table.~\ref{tab_table1}.

Our results hint at the existence of additional microscopic conserved (or quasi-conserved) quantities that give rise to non-ergodic behaviour in a quadratic trap even when the Lyapunov exponents are positive. The special case of $N=3$ displays non-chaotic (zero Lyapunov exponent) behaviour. On the other hand, hard rods confined to quartic traps exhibit conventional non-integrable behaviour, namely positive Lyapunov exponents and thermalization to the expected Gibbs state.

\begin{table}[b]
	\centering
	\begin{tabular}{|c|c||c|}
		\hline $N>3$
		& Quadratic &  Quartic \\ 
		\hline
		Chaos &  Yes & Yes   \\ 
		~& (Fig.~\ref{fig:N4timeseries} and~\ref{fig:LLExpc}) & (Fig.~\ref{fig:quartproblambda} and~\ref{fig:LLExpc}) \\ 
		\hline
		Ergodicity & No &  Consistent with yes \\ 
		~ & (Fig.~\ref{fig:harmproblambda}) &  (Fig.~\ref{fig:quartproblambda}) \\ 
		\hline
		Thermalization & No& Yes \\ 
		~& (Fig.~\ref{fig:harmNCTherm} and \ref{fig:harmNonGibbs}) & (Fig.~\ref{fig:quartNCTherm}) \\ 
		\hline
	\end{tabular}
	\caption{The table provides a summary of our findings for $N>3$ hard rods confined to quadratic and quartic traps. The case of $N=3$ rods in a quadratic trap is special (Fig.~\ref{fig:LLExpN3}a,b) because it is characterised by vanishing Lyapunov exponents although its quartic counterpart, even for $N=2$, has non-zero Lyapunov exponents (Fig.~\ref{fig:LLExpN3}c,d). }
	\label{tab_table1}
\end{table}

Our work suggests several interesting open questions for hard rods in a quadratic trap such as: (i) finding the extra conservation law for $N=3$, assuming this exists (it was previously argued that any such conservation law must be non-analytic in the dynamical variables~\cite{cao2018incomplete}); (ii) understanding the dependence of $\lambda$ on energy ${\rm e}$ and $N$ (see Fig.~\ref{fig:LLExpc}) ;  (iii) understanding whether hydrodynamics can capture the regime of intermediate times between the initial and late-time dynamics~\cite{cao2018incomplete}; (iv) exploring whether this lack of ergodicity for large $N$ has any relation to the known additional, ``entropic'' conservation laws of ballistic-scale GHD~\cite{doyon2017note,bulchandani2017classical,cao2018incomplete,caux2019hydrodynamics}, or some hitherto undiscovered conservation laws of the full dissipative hydrodynamics.

We expect that some of our findings will be valid more generally for systems of classical or quantum particles confined to a trap that breaks the integrability of their interactions. We note that studies of the Toda chain~\cite{di2018transport,dhar2019transport} have also indicated drastic differences in transport properties in a quadratic trap compared to quartic traps. As a more extreme example of such unusual behaviour, the rational Calogero model remains integrable in both quadratic and quartic traps~\cite{Polychronakos_2006}, and its ballistic scale hydrodynamics is integrable in any trap~\cite{bulchandani2021quasiparticle}. A complete theory of this rich phenomenology of integrability breaking by traps remains elusive for now.

\section{Acknowledgements}
M.K. would like to acknowledge support from the project 6004-1 of the Indo-French Centre for the Promotion of Advanced Research (IFCPAR), Ramanujan Fellowship (SB/S2/RJN-114/2016), SERB Early Career Research Award (ECR/2018/002085) and SERB Matrics Grant (MTR/2019/001101) from the Science and Engineering Research Board (SERB), Department of Science and Technology (DST), Government of India. A.K. acknowledges the support of the core research grant CRG/2021/002455 and the MATRICS grant MTR/2021/000350 from the SERB, DST, Government of India. A.D., M.K., and A.K. acknowledge support of the Department of Atomic Energy, Government of India, under Project No. 19P1112RD.  D.A.H. was supported in part by (U.S.A.) NSF QLCI grant OMA-2120757. We would like to acknowledge  the ICTS program - "Hydrodynamics and fluctuations - microscopic approaches in condensed matter systems (code: ICTS/hydro2021/9)" for enabling discussions. A.D. and M.K. acknowledge the support from the Science andEngineering Research Board (SERB, government of India),under the VAJRA faculty scheme (No. VJR/2019/000079).


\bibliography{References.bib}

\end{document}